\documentclass[twocolumn]{article}

\usepackage{geometry}
\usepackage{cite}
\usepackage{caption}
\usepackage{graphicx}
\usepackage{amsmath}
\usepackage{amssymb}
\usepackage{textcomp}
\usepackage{lmodern}

\newcommand{\onlinecite}[1]{\hspace{-1 ex} \nocite{#1}\citenum{#1}} 

\begin{document}

\twocolumn[
\begin{@twocolumnfalse}
\begin{center}
\LARGE Optoelectronic Intelligence\\
\vspace{0.2em}
\Large Jeffrey M. Shainline\\
\vspace{0.0em}
\textit{\small National Institute of Standards and Technology, Boulder, CO, 80305}\\
\vspace{0.3em}
\small October 16th, 2020
\end{center}
\begin{abstract}
To design and construct hardware for general intelligence, we must consider principles of both neuroscience and very-large-scale integration. For large neural systems capable of general intelligence, the attributes of photonics for communication and electronics for computation are complementary and interdependent. Using light for communication enables high fan-out as well as low-latency signaling across large systems with no traffic-dependent bottlenecks. For computation, the inherent nonlinearities, high speed, and low power consumption of Josephson circuits are conducive to complex neural functions. Operation at 4\,K enables the use of single-photon detectors and silicon light sources, two features that lead to efficiency and economical scalability. Here I sketch a concept for optoelectronic hardware, beginning with synaptic circuits, continuing through wafer-scale integration, and extending to systems interconnected with fiber-optic white matter, potentially at the scale of the human brain and beyond.
\vspace{3em}
\end{abstract}
\end{@twocolumnfalse}
]

\section{\label{sec:introduction}Introduction}
General intelligence is the ability to assimilate knowledge across content categories and to use that information to form a coherent representation of the world. The brain accomplishes general intelligence through many specialized processors performing unique, complex computations \cite{ba1988,de2014}. The information generated by these processors is communicated throughout the network via dedicated connections spanning local, regional, and global scales \cite{sp2010}. On the micro-scale, synapses, dendrites, and neurons are specialized processors comprising the gray matter computational infrastructure of the brain \cite{geki2002}. On the meso-scale, cortical minicolumns of 100 neurons act as specialized processors \cite{mo1997}, and on the macro-scale, brain regions play that role \cite{brme2010}. Information is communicated between minicolumns and between brain regions via axonal fibers that comprise the white matter communication infrastructure of the brain. On short time scales, information processing occurs in synapses \cite{abre2004}, dendrites \cite{stsp2015}, and within single neurons \cite{ko1997}. On longer time scales, the information generated by minicolumns is communicated across wider regions of the network so that the knowledge of specialized processors can combine in a comprehensive interpretation of a subject \cite{bu2006}. The utilization of many specialized processors combining their shared computational resources across many scales of space and time enables the brain to achieve general intelligence \cite{ba1988,de2014}. 

Computation and communication are the complimentary pillars of neural systems. Hardware for artificial general intelligence (AGI) will achieve the highest performance if complex, local processors can pool the information from their specialized computations through global communication. Electrons excel at computation, while light is excellent for communication. In silicon hardware, monolithic optical links between a processor and memory have been demonstrated \cite{suwa2015}. These devices were fabricated in a 45-nm CMOS node with no in-line process changes, and off-chip light sources were utilized. Such work is driven by the desire for increased communication bandwidth in multi-core architectures. These architectures continue to expand into on-chip networks, in some cases resulting in highly distributed, brain-inspired systems implemented with CMOS electronics \cite{bo2000,pfgr2013,mear2014,fuga2014,payu2017,dasr2018}. As computing grows more distributed, communication becomes a bottleneck. A primary challenge affecting further chip-scale electronic-photonic integration is the difficulty of achieving a light source on silicon that is robust, efficient, and economical \cite{libo2010,zhyi2015}. Lessons learned from very-large-scale integration (VLSI) inform us that economical fabrication of integrated circuits comprising simple components is necessary for scaling. In this regard, difficulties associated with integrated light sources are the most significant impediment to optoelectronic VLSI.

\begin{figure*}[t!]
    \centering{\includegraphics[width=17.6cm]{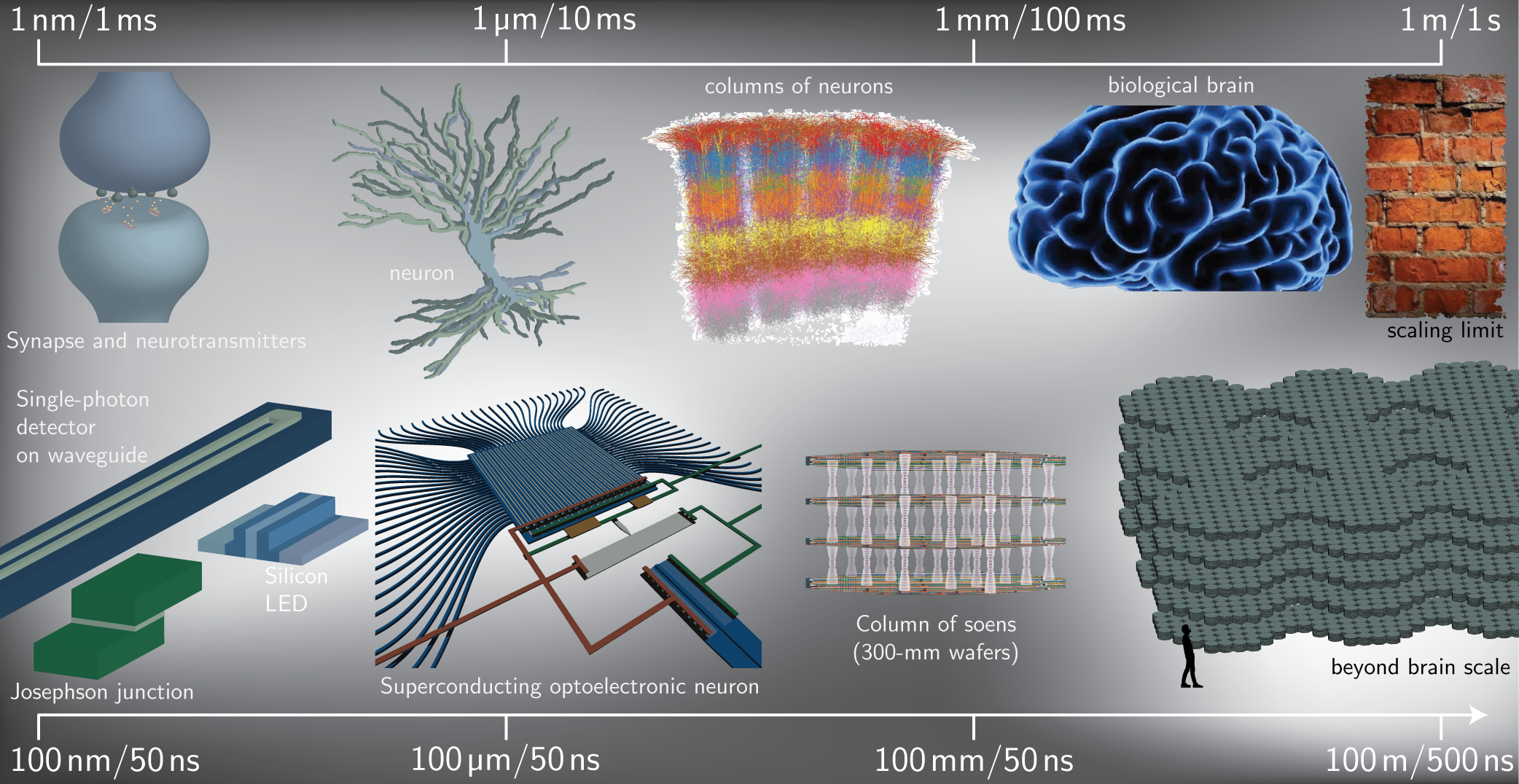}}
	\captionof{figure}{\label{fig:complexityAcrossScales}Structure across scales. Biological systems have functional components from the nanometer scale (synapses, neurotransmitters and receptors, axonal pores, etc.) up to the full brain (roughly 0.3\,m linear dimension for full human cerebral cortex \cite{sh2018_ICRC}). The speeds of various operations are limited by chemical diffusion and signal propagation along axons, which may ultimately limit the size of biological neural systems \cite{bu2006,sh2018_ICRC}. The time constants associated with chemical diffusion and membrane charging/discharging span the range from 1\,ms to 100\,ms \cite{geki2002} and significantly influence the speeds at which information processing occur. The wall at the right indicates the communication-limited spatial-scaling barrier. By contrast, optoelectronic devices rarely have functional components with critical dimension smaller than 100\,nm, and optoelectronic neurons are likely to be on the 100-\textmu m scale, with dendritic arbor extending for millimeters and axonal arbor in some cases spanning the system. The time constants of these components can be engineered in hardware across a very broad range with high accuracy through circuit parameters, enabling rapid processing as well as long-term signal storage.  Yet optical communication between wafers enables vertical stacking into columns, and the high speed of devices and communication enables optoelectronic systems to extend far beyond the limits imposed by the slow conduction velocity of axons. In the large system illustrated at the right, photonic communication would enable every neuron of the system to simultaneously fire at its maximal rate with no traffic-induced communication delays.}
\end{figure*}

It is the perspective of our group at NIST that hardware incorporating light for communication between electronic computational elements combined with an architecture of distributed optoelectronic spiking neurons will provide potential for AGI at the scale of the human brain. While much of present-day computing infrastructure has evolved to implement a von Neumann architecture perfoming sequential operations in the model of a Turing machine, the functioning of neural systems departs considerably from this model. Light has even more to offer in a neural computing context, because communication across scales is indispensable. Further, the spiking behavior of Josephson junctions combined with the efficiency of single-photon detectors make a compelling case for optical integration with superconducting electronics \cite{shbu2017,sh2018}. Such a choice necessitates low-temperature operation near 4\,K. At this temperature, silicon light sources become available \cite{buch2017}, indicating that a major impediment to optoelectronic VLSI many not be present in the superconducting domain. This article summarizes the reasoning behind the assertion that superconducting optoelectronic systems have unique potential to achieve general intelligence when considered from the perspectives of cognitive science and VLSI.

\section{\label{sec:neuralArchitectureAndDevices}Neural devices and architecture}
To guide the design of hardware for AGI, we must consider operation across spatial and temporal scales \cite{beba2016}. Figure \ref{fig:complexityAcrossScales} charts the structures present on various scales for biological and optoelectronic hardware. The human brain has features spanning roughly eight orders of magnitude in size, from a nanometer to a tenth of a meter. Across time, activity ranges from the 1\,ms time scale of neurotransmitter diffusion across a synapse, through the 200\,ms time scale of brain-wide theta oscillations, up to the memory retention time of the organism. The speeds of devices and communication in the brain are limited by the chemical and ionic nature of various operations. The maximum size of the brain may be limited by the slow conduction velocity of ionic signals along axons. If the brain were larger, signals would not have time to propagate between different regions during the period of theta oscillations, and system-wide information integration could not be efficiently achieved \cite{bu2006,sh2018_ICRC}. 

Light and electronics together can enable communication and computation across spatial and temporal scales. We have proposed a specific approach we see as most conducive to large-scale implementation for AGI \cite{shbu2017,sh2018,sh2018_ICRC,sh2019_fluxonic}. The approach combines waveguide-integrated light sources and single-photon detectors for communication \cite{shbu2017,buch2017} with Josephson circuits for synaptic, dendritic, and neuronal computation \cite{sh2018,sh2019_fluxonic}. As illustrated in Fig.\,\ref{fig:complexityAcrossScales}, these optoelectronic networks will have features as small as 100\,nm and potentially extend up to kilometers. Neuronal inter-spike intervals can be as short as 50\,ns, while synaptic and dendritic processing occurs on the 50\,ps time scale of Josephson junctions. Time constants can be chosen across many orders of magnitude, enabling information processing and memory across time scales. Figure\,\ref{fig:complexityAcrossScales} is intended to emphasize that if communication barriers can be removed, neural systems of extraordinary scale can be achieved.

\begin{figure}[tb]
	\centering{\includegraphics[width=8.6cm]{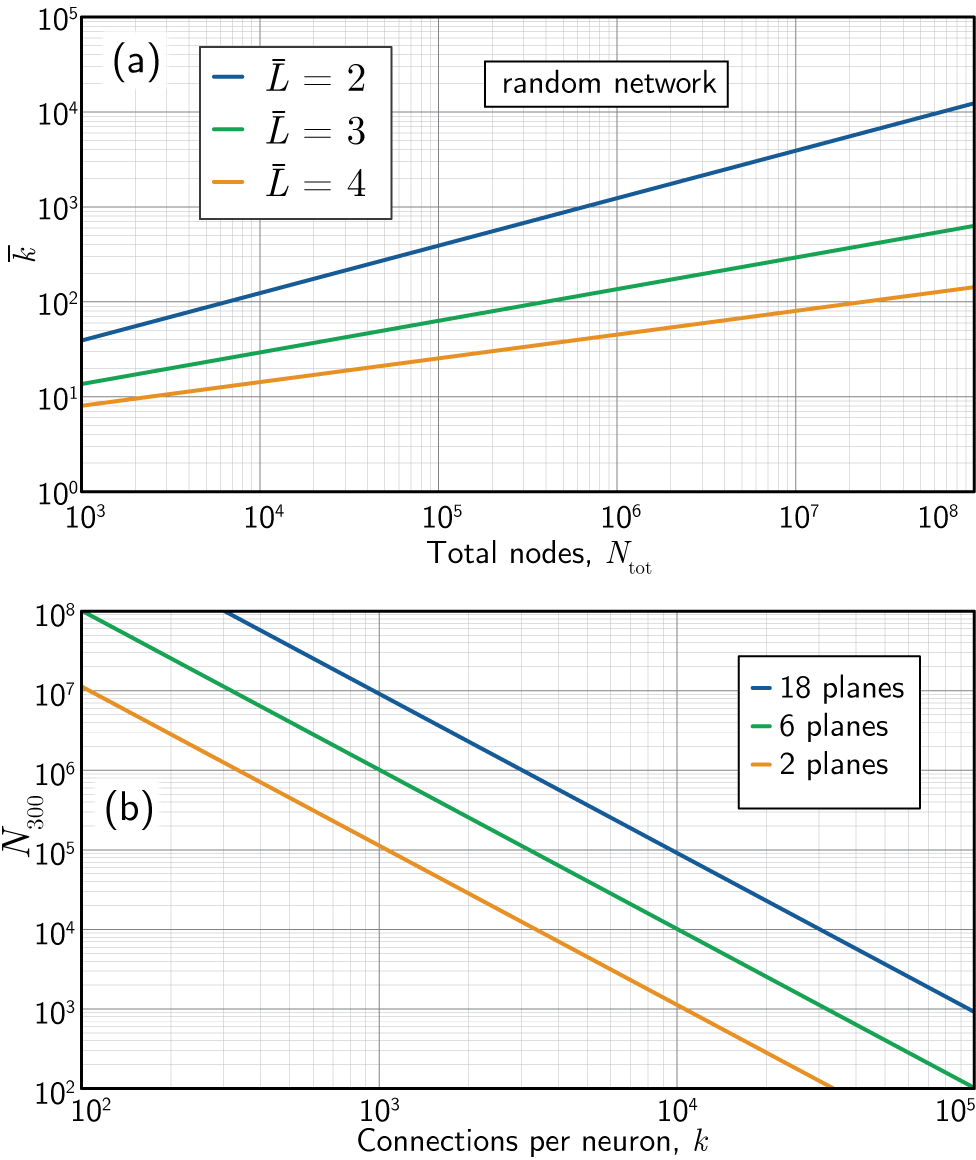}}
	\captionof{figure}{\label{fig:data}Scaling considerations for optoelectronic neural systems. (a) The average number of connections per node ($\bar{k}$) required to maintain a given average path length ($\bar{L}$)  across a random network as a function of the total number of nodes in the system ($N_{\mathrm{tot}}$). (b) The total number of nodes that can fit on a 300\,mm ($N_{300}$) wafer as a function of the number of connections per node ($k$) for various numbers of waveguide planes in the wire-limited regime \cite{ke1982}.}
\end{figure}

Information processing in neural systems employs clusters of neurons to represent specific features, and the information from these clusters must be shared with other regions of the network to form a multifaceted representation of a stimulus. Structurally, this information processing is facilitated by networks with a high clustering coefficient yet also an average path length nearly as short as a random graph \cite{eskn2015}. Such small-world networks \cite{wast1998} are ubiquitous throughout the brain \cite{sp2010} and require long-range connections. In a random network, near and distant connections are equally probable, so the average path length across the network achieves a lower limit on path length for a given number of edges connecting a given number of nodes. Figure \ref{fig:data}(a) shows the number of edges required per node to achieve a given average path length as a function of the number of nodes in the random network. For a modest network with one million nodes, each node must make one thousand connections to maintain a path length of two. For the case of a network with 100 million nodes, each node must make 10,000 connections. This is similar to the hippocampus in the human brain, with nearly 100 million neurons, each with 10,000 or more nearly random synaptic connections \cite{bu2006}. Maintaining a short path length across the network is critical for information integration, and is an important motivator to use light for communication.

In addition to small-world characteristics, networks of the brain demonstrate a hierarchical architecture wherein minicolumns aggregate into columns, columns into modules, and modules into complexes. This fractal property is necessary to enable networks to scale arbitrarily, with dynamics constrained only by the physical hardware and spatial extent of the system rather than by the ability to communicate across the network \cite{plth2006}.  Hierarchical architecture with simple device behavior can generate some of the dynamical features of the brain, such as neuronal avalanches \cite{plth2006} that are observed in the resting state \cite{peth2009}. Other dynamical behaviors thought to be necessary for attention, cognition, and learning, such as cross-frequency coupling \cite{bu2006} and synaptic plasticity \cite{mage2012,ab2008,fudr2005}, require complex capabilities at the device level. Dynamical synapses, dendrites, and neurons allow one structural network to realize myriad functional networks adapting on multiple time scales. While light is excellent for communication, electrical circuits are better equipped to perform these nonlinear, dynamical functions. In particular, Josephson circuits naturally manifest many of the neuromorphic operations we seek \cite{sh2018,sh2019_fluxonic}. For communication and computation, neural information processing will benefit immensely from optoelectronic integration.

\section{\label{sec:synapsesDendritesAndNeurons}Optoelectronic synapses, dendrites, and neurons}
Having chosen to communicate synaptic events with light, the quantum limit is a single photon per synaptic connection. We have designed a synapse \cite{sh2018,sh2019_fluxonic} that detects a single near-infrared photon and requires no power to retain the synaptic state, a feature enabled by the dissipationless nature of superconductors. The synapse utilizes a superconducting-nanowire single-photon detector (SPD), which is simply a current-biased strip of superconducting wire \cite{mave2013}. To achieve the desired synaptic operation, an SPD is combined in circuits with Josephson junctions (JJs) and superconducting loops to achieve the functions needed for neural information processing. In optoelectronic synapses of this design, the current bias across a single JJ establishes the synaptic weight. This current bias can be dynamically modified through various photonic and electronic means based on control signals or network activity. Dendritic and neuronal nonlinearities are a natural consequence of the JJ critical current, and these functions were also explored in Refs. \onlinecite{sh2018} and \onlinecite{sh2019_fluxonic}. Due to the prominent role of superconducting current storage loops, we refer to these as loop neurons. We refer to networks of loop neurons as superconducting optoelectronic networks (SOENs). In the operation of loop neurons, a single photon triggers a synaptic event, and spike-timing-dependent plasticity is induced by two photons\textemdash one from each neuron associated with the synapse.

In addition to the choice of SPDs as the detectors in the system, we must also select a light source, which must be fabricated across wafers by the millions. Because our choice of detectors dictates cryogenic operation, silicon light sources are an option. The light sources we have in mind are silicon LEDs \cite{buch2017}, employing luminescence from defect-based dipole emitters. From the perspective of VLSI, achievement of a silicon light source as simple as a transistor would be the greatest contribution to the success of this technology. If cryogenic operation enables both single-photon detectors and silicon light sources, it will be worth the added infrastructure for cooling.

\begin{figure*}[tb]
    \centering{\includegraphics[width=17.2cm]{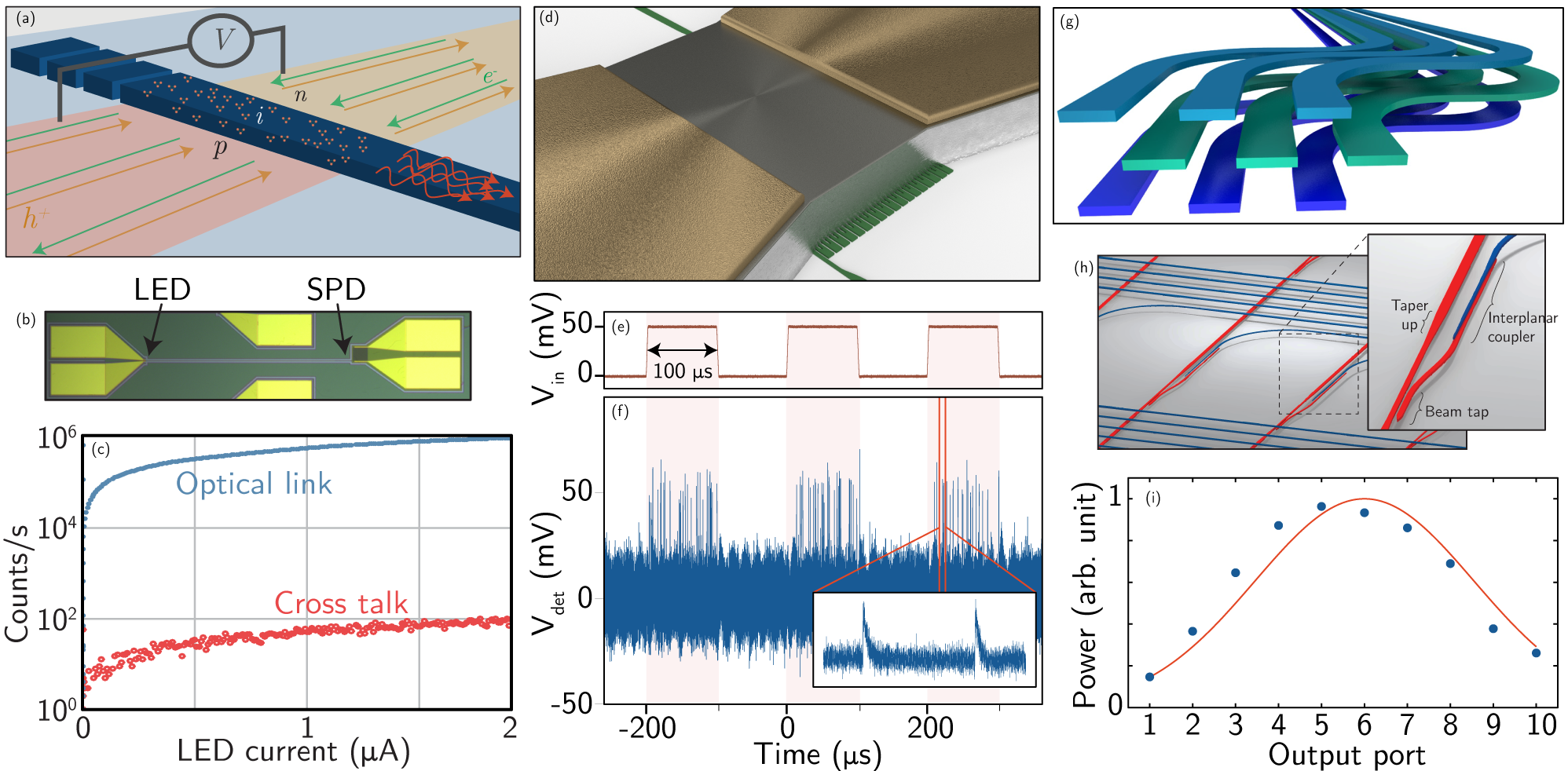}}
	\captionof{figure}{\label{fig:experiment}Experimental progress toward superconducting optoelectronic networks. (a) Schematic of waveguide-integrated silicon LED. Embedded emitters are shown in the intrinsic region of the $p-i-n$ junction. (b) Microscope image of a silicon LED waveguide-coupled to a superconducting-nanowire detector. (c) Experimental data showing that light is coupled through the waveguide, while cross talk to an adjacent detector on the chip is suppressed by 40\,dB. (a-c) Adapted from Ref.\,\onlinecite{buch2017}. (d) Schematic of the superconducting thin-film amplifier. The superconducting channel is shown in green. The resistive gate is grey with gold pads. When current is driven through the gate, superconductivity is broken in the channel, leading to high impedance of $\approx 1$\,M$\Omega$. (e,f) The resistive switch driving the LED. (e) Square pulses are driven into the switch gate. (f) When the switch is driven, light is produced from the LED and detected by the SPD. In (e,f), the switch was on one chip, and the light source and detector were on a different chip. The same light source and detector of (a-c) were employed in this demonstration. (d-f) Adapted from Ref.\,\onlinecite{mcve2019}. (g) Schematic of multi-planar integrated waveguides for dense routing. (h) Schematic of feed-forward network implemented with two planes of waveguides. The inset shows the elemental tap and transition device. (i) Data from an experimental demonstration of routing between nodes of a two-layer feed-forward network with all-to-all connectivity. The data is from light at a single input and collected at all ten outputs with the designed Gaussian distribution profile. (g-i) Adapted from Refs.\,\onlinecite{chbu2017} and \onlinecite{chbu2018}.}
\end{figure*}
To achieve complex neural circuits, we aim for monolithic integration of light sources, detectors, and superconducting circuit elements. Our group's experimental progress towards this end is summarized in Fig.\,\ref{fig:experiment}. We demonstrated waveguide coupling of light from micron-scale, all-silicon LEDs to integrated SPDs on a silicon photonic chip (Fig.\,\ref{fig:experiment}(a-c), \cite{buch2017}). The performance achieved in the first iteration of these optical links was not yet adequate. The observed efficiency was $5\times10^{-7}$, while $10^{-3}$ is desirable for large systems \cite{sh2018}. Yet the simplicity of both the source and detector made the fabrication and demonstration of a monolithic optical link far easier than if room-temperature operation were required. We also recently demonstrated superconducting amplifiers driving these semiconductor light sources \cite{mcve2019}. For this application, the light sources are only required to produce incoherent pulses of 10,000 photons ($\approx 1$\,fJ) at 20\,MHz when operating at 4\,K. If silicon light sources can meet these performance specifications, the hardware stands a chance of enabling brain-scale systems with 30,000 times the speed.

We have also demonstrated superconducting amplifiers capable of generating the voltage required to produce light from these sources (Fig.\,\ref{fig:experiment}(d-f), \cite{mcve2019}). Generating more than a millivolt with superconducting circuits is difficult, but the thin-film, micron-scale cryotron demonstrated in Ref.\,\cite{mcve2019} leverages the extreme nonlinearity of the superconducting phase transition to rapidly generate high impedance and voltage with low energy, thus driving a semiconductor light source during each neuronal firing event. In Ref.\,\onlinecite{mcve2019} we demonstrate the use of these amplifiers to drive the LED-SPD link of Ref.\,\onlinecite{buch2017}.

During neural operation, the light produced by these LEDs fans out across a network of micron-scale dielectric waveguides terminating on the superconducting detectors at each synaptic connection. We have demonstrated multiple vertically integrated planes of these waveguides (Fig.\,\ref{fig:experiment}(d-f), \cite{chbu2017}), and used them to implement the architecture of a feed-forward neural network with two layers of 10 neurons per layer and all-to-all connectivity \cite{chbu2018}.

This approach to neural computing resides at the confluence of superconducting electronics and integrated photonics, and therefore must be contrasted with other work in these fields. Regarding photonic neural systems, excitable lasers have been explored as spiking neurons (see Ref.\,\onlinecite{prsh2017} and references therein). These lasers integrate several optical inputs, and release a laser pulse upon reaching threshold. These devices can be extremely fast, but consume too much power for scaling to the level of the human brain. Excitable lasers can be used as spiking neurons in the broadcast-and-weight architecture \cite{tana20142}, wherein each neuron is assigned a wavelength, and synaptic weights are established with microring resonators that attenuate the optical signals, much like  waveglength-division-multiplexed (WDM) fiber-optic networks. In conventional silicon photonics, WDM employs on the order of 10 channels (1\,THz divided by 100\,GHz). It may be possible to extend this to order 100 for highly connected neural systems \cite{tana20142}, but such dense channel packing would require cumbersome control to hold stable synaptic weights without cross talk. The requirement of precise control at every synapse as well as the non-monotonic, rapidly varying Lorentizian lineshape of microring resonances precludes unsupervised learning. 

Phase change materials have also been explored for neuronal thresholding \cite{chsa2018} and as a means of implementing variable attenuation of photonic signals to establish a synaptic weight \cite{chri2017}. This approach requires billions of photons to achieve synaptic weight modification, and relying on the properties of a material to achieve the complex computations occurring at a synapse results in limited functionality as compared to optoelectronic circuits. Deep learning with continuous fields rather than spiking neurons is also receiving attention, and networks of on-chip, cascaded Mach-Zehnder interferometers are a prominent approach \cite{shha2016}. Such networks perform matrix-vector multiplication, but are not conducive to establishing the recurrent networks employed by spiking neural systems nor the activity-dependent plasticity necessary for unsupervised learning. 

To our knowledge, all other proposed photonic neural systems encode information in the amplitude of optical signals, and synaptic weights are established through attenuation of these signals. We use light for binary communication, establishing synaptic weights through electronic responses that consume less energy, thereby keeping optical power levels at an absolute minimum. While operation at 4\,K brings a factor of one thousand power penalty for cooling, waveguide-integrated, room-temperature photodetectors require several thousand photons to register an event. Thus, the power penalty of cryogenic operation is compensated by the ability to detect single photons, even before the efficiency gains of cold light sources and superconducting electronics are considered.

Approaches to neural computing using superconducting circuits leverage the nonlinear, spiking properties of Josephson junctions \cite{hias2007,sele2017,scdo2018}. The most successful experimental effort to date demonstrated coupling of two neurons based on JJs, with inter-spike intervals on the order of tens of picoseconds \cite{sele2017}. The challenge with using superconducting electronics is communication. In superconducting circuits, direct fan-out is usually limited to two, so for neurons to make thousands of connections, many stages of pulse splitters and active transmission lines must be employed. This leads to a cumbersome communication network requiring many JJs and severe challenges for routing. 

To our knowledge, all other photonic and superconducting electronic neuromorphic efforts are focused on attaining relatively small-scale systems to accomplish specific computational functions\textemdash programmable accelerators or experimental test beds rather than cognitive systems. The superconducting optoelectronic approach championed here utilizes similar superconducting circuits as Refs.\,\onlinecite{hias2007,sele2017,scdo2018} for synaptic, dendritic, and neuronal computation, while leveraging light for communication, thereby enabling scaling to massively interconnected systems. Electrons and photons are complimentary, and optoelectronic integration is most straightforward when combining superconducting circuits with silicon light sources operating at liquid helium temperature.

\section{\label{sec:communication}Scaling an optoelectronic system}
To further explain why we place optical communication at the center of hardware development, I briefly summarize the physical limitations of electrical interconnection networks \cite{hepa2012}. It is impracticable in silicon electronics for a single device to source current to many other devices. A shared communication network must be employed. Switched media networks are used for this purpose. Each device must then only communicate to the nearest switch in the network. Because the communication infrastructure is shared, devices must request and wait for access to the switch network to transmit messages. This approach to communication leverages the speed of electronic circuits to compensate for the challenge of direct communication. Limitations are reached when many devices must communicate with many other devices simultaneously. While neural activity is generally sparse, during neuronal avalanches, or during coordinated gamma bursting, many neurons must communicate simultaneously across the network. As more neurons, each with many synapses, are added to the network, the average frequency of neuronal firing events must decrease. Integration of information across the network is limited by the communication infrastructure.

The physics of light is complementary to that of electrons. Photons can co-propagate on a waveguide independently without capacitance. Waveguides can fan out without a charging penalty due to wiring. This is not to say photonic communication can address an arbitrarily large number of recipients without consequence. For each new recipient, the number of photons in a neuronal pulse must increase. As destinations get further away, more energy is dissipated to propagation loss. These realities notwithstanding, it is feasible for devices communicating with photons to make direct, independent connections to thousands of destinations, thereby eliminating the need for the shared communication infrastructure that is the primary impediment to achieving AGI with electrical interconnections.

Having made this claim, the burden is upon us to provide evidence of the feasibility of photonic communication in large-scale neural systems. The large wavelength of light relative to the size of electronic devices causes concern for the size of optoelectronic brain-scale networks. To build confidence for the feasibility of the endeavor, I sketch here a vision of how such an optoelectronic neural system may be constructed. At the foundation of this vision is the assumption that the technology will utilize the fabrication infrastructure of silicon electronics and photonics in conjunction with fiber optics for longer-range communication. 

\begin{figure*}[tb]
    \centering{\includegraphics[width=17.6cm]{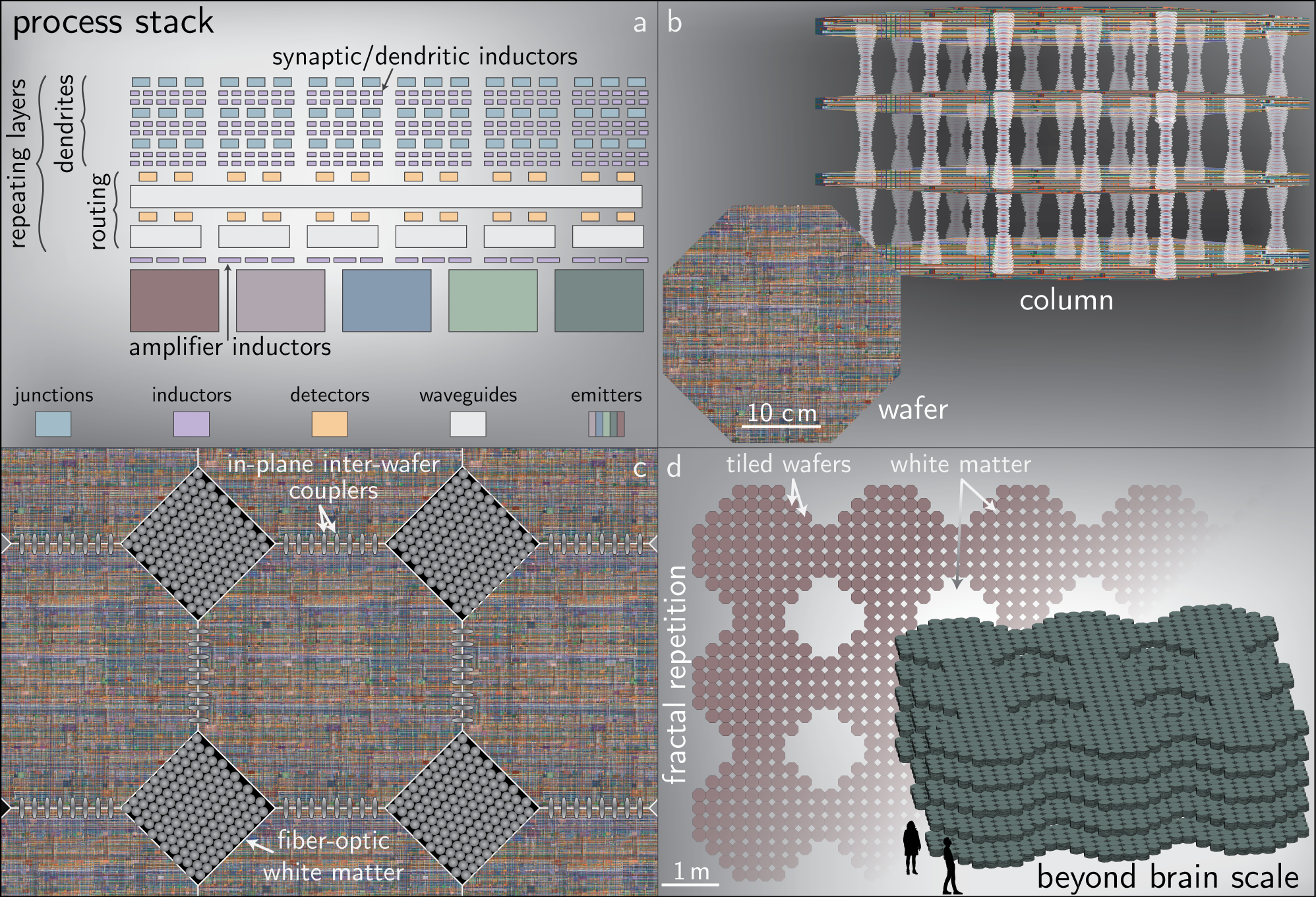}}
	\captionof{figure}{\label{fig:communicationAcrossScales}Hierarchical construction of optoelectronic neural systems. (a) Schematic of the process stack, with silicon light sources on a silicon-on-insulator wafer, waveguides and detectors above, followed by the Josephson infrastructure and mutual inductors for dendritic processing. (b) Vertical photonic communication between two stacked wafers. Liquid helium flows between the wafers of a column for cooling, and free-space links propagate without loss through the helium. The inset shows a schematic of a single 300\,mm wafer, with neurons and routing, cut into an octagon for tiling. (c) Illustration of in-plane tiling. Lateral wafer-edge links connect wafers in a plane, and fiber optic bundles fill the voids between wafers for long-range communication. (d) A large neural system with multiple large modules, each containing hundred to thousands of wafers, enabled by photonic communication and the efficiency of superconducting detectors and electronics. Not shown is the fiber-optic white matter that would be woven through the voids between the octagons in this example hierarchical tiling.}
\end{figure*}

At the wafer scale, light will be guided in multiple planes of dielectric waveguides \cite{chbu2017,chbu2018} (Fig.\,\ref{fig:communicationAcrossScales}(a)), just as integrated electronics requires multiple wiring layers. To estimate the area of such photonic interconnection networks, we follow Keyes \cite{ke1982} and approximate the number of neurons that can be supported on a 300-mm wafer by $N = 2\sqrt{2}r^2\left(p/wk_{\mathrm{in}}\right)^2$. Here, $p$ is the number of planes of waveguides, $w$ is the waveguide pitch (1.5\,\textmu m), $k_{\mathrm{in}}$ is the number of waveguides entering the neuron, and $r = 150$\,mm. The prefactor results from assuming octagonal tiling. This expression is plotted in Fig.\,\ref{fig:data}(b). The estimate informs us that a 300-mm wafer with six waveguide planes can support roughly one million neurons if they each have one thousand connections. More involved analysis finds more planes may be needed \cite{sh2018_ICRC}. As a point of comparison to electrical neural systems, Ref.\,\onlinecite{kuwa2017} finds that through multi-layer, wafer-scale integration of logic and memory, 250 million electrical neurons could fit on a 300\,mm wafer. The trade-off is speed, as the shared communication network would limit the electrical neurons studied in Ref.\,\onlinecite{kuwa2017} to 10\,Hz operation when 1000 synaptic connections are made per neuron. Nevertheless, the message of Fig.\,\ref{fig:data}(b) is that photonic routing results in large area consumption. An optoelectronic brain larger than that of a bumble bee will not fit on a single 300-mm wafer.

Optoelectronic intelligence will require communication between wafers. Wafers can be stacked vertically, and free-space optical links can send photons from a source on one wafer to a detector on a wafer above or below \cite{caga2012}, as illustrated in Fig.\,\ref{fig:communicationAcrossScales}(b). Assuming SPDs receiving vertical communication have a pitch of 25\,\textmu m, a 300-mm octagon could support $10^8$ vertical communication links between two wafers. Considering wafers as laminar layers, as in cortex, such a configuration would result in roughly 5\% inter-layer connectivity, similar to the fraction observed in mammals (Ref.\,\onlinecite{bu2006}, pg. 286).

In addition to feed-forward and feed-back free-space vertical coupling, lateral inter-wafer communication can be achieved at wafer edges, as shown in Fig.\,\ref{fig:communicationAcrossScales}(c). In the tiling considered here, each wafer makes such connections to neighbors in the cardinal directions. With a 10\,\textmu m pitch, 11,500 wafer-edge couplers could be supported in each direction. Such a system would demonstrate strong connectivity within the vertical stack of the wafers, and weaker lateral connectivity. The reader may recognize the columnar organization of the cerebral cortex \cite{mo1997}.

To achieve communication from within these columns to other regions of the network, optical fibers are ideal. Within the tiling under consideration, the square areas at diagonals between wafers can support fiber-optic bundles (Fig.\,\ref{fig:communicationAcrossScales}(c)). These optical fiber tracts are analogous to white matter in the brain. One such region could house a million single-mode fibers of 125\,\textmu m diameter. These fibers will emanate from all wafers within the column, and if six wafers are stacked in a column, each wafer would have 167,000 output fibers to carry information to other regions. With one million neurons on a wafer, not every neuron would have access to a fiber for long-distance communication (unless wavelength multiplexing is employed). This again is consistent with brain organization, wherein the number of long-distance axons emanating from a region is smaller than the number of neurons within the region. Each of these fibers can branch as it extends through the white matter, so a neuron with access to a single wafer-edge fiber can establish multiple long-range connections. 

With this columnar configuration in mind, one can assess the feasibility of constructing a system on the scale of the human cerebral cortex (10 billion neurons, each with thousands of synaptic connections). If a wafer holds a million neurons, a cortex-scale assembly requires 10,000 wafers. Assuming the volume of white matter scales as the volume of grey matter to the 4/3 power \cite{zhse2000}, the cortex-scale system would fit in a volume two meters on a side. While optoelectronic neurons are significantly bigger than their biological counterparts, it is not the absolute size that limits system performance. The relevant quantity for assessing scaling limitations is the size of a neuron divided by the velocity of communication \cite{sh2018_ICRC}. Communication at the highest velocity in the universe more than compensates. 

Regarding power, a single 300-mm wafer with a million neurons would dissipate one watt if the light production efficiency were $\eta = 10^{-4}$, a conservative estimate. For the cortex-scale system of 10,000 wafers, the device power consumption with $\eta = 10^{-4}$ would be 10\,kW. A further power penalty of one thousand would be incurred if the system were operated in a background of 300\,K. Thus, even in a conservative case of poor light production efficiency, an AGI on the scale of the human brain would consume 10\,MW, the same order as a modern supercomputer. We are considering a system with roughly the same number of neurons and synapses as the human cerebral cortex, but with activity at 30,000 times the speed. While there is high uncertainty associated with scaling estimates of such an immature technology, these calculations indicate that artificial brain-scale systems with photonic communication and electronic computation may be feasible, a possibility with profound implications for the future of science and technology. 

\section{\label{sec:discussion}Summary and Discussion}
I have argued that artificial neural hardware should be designed and constructed to leverage photonic communication while performing synaptic, dendritic, and neuronal functions with electronic devices. Superconducting optoelectronic circuits elegantly implement these functions, in part because of the utility of Josephson nonlinearities for neural computation, and also because superconducting detectors enable few-photon signals, approaching the lowest possible energy for optical communication. We have demonstrated all of the core components and are working toward complete integration.

This approach to AGI appears possible for physical and practical reasons. Physically, due to photonic signaling, it is possible to achieve efficient communication across the network for systems with orders of magnitude more than the 10,000 wafers comprising a brain-scale system. Reference\,\onlinecite{sh2018_ICRC} explores the communication-limited size of the system as a function of the frequency of network oscillations. Specialized processors with activity at 20\,MHz (the gamma firing rate of loop neurons) can span an area 10 meters on a side before delays limit communication. Modules with activity at 1\,MHz (the frequency of corresponding theta oscillations in this system) could integrate information across an area the size of a data center within a single theta cycle.

On the practical side, fabrication of SOENs at industrial scale appears feasible. All the proposed circuits can be created on 300-mm wafers with existing infrastructure, such as a 45-nm CMOS node. Ten thousand wafers move through such a foundry every day. If dedicated to fabrication of optoelectronic intelligence, a foundry could produce multiple brain-scale systems per year. While the devices employed here depart from conventional silicon microelectronics, the same fabrication infrastructure can be employed. 

What are the next steps to realize loop neurons and SOENs? Low-cost source-detector integration at the wafer scale is required. Active devices must be augmented with improvements in deposited dielectrics to enable many planes of routing waveguides with low loss. Hardware improvements will not lead to AGI without further theoretical insights. Conceptual advances are required to achieve high-performance neural systems, train them, and make them intelligent.

\vspace{2em}
This is a contribution of NIST, an agency of the US government, not subject to copyright.

\section{\label{sec:acknowledgements}Acknowledgements}
I acknowledge significant contributions to this project from team members Dr. Sonia Buckley, Dr. Jeff Chiles, Dr. Saeed Khan, Dr. Adam McCaughan, and Dr. Alexander Tait. This work would not be possible without the group leadership of Dr. Sae Woo Nam and Dr. Richard Mirin and the institutional support of NIST.

\bibliographystyle{unsrt}
\bibliography{optoelectronic_intelligence}

\end{document}